\documentclass[onecolumn,preprint,preprintnumbers,amsmath,amssymb]{revtex4}

\usepackage{graphicx}
\usepackage{dcolumn}
\usepackage{bm}

\begin{document}

\title{General method for focusing of waves using phase and amplitude compensation}

\author{L.E. Helseth}

\affiliation{Department of Physics and Technology, University of Bergen, N-5007 Bergen, Norway}%

\begin{abstract}
A general method for focusing of waves, based on phase and amplitude compensation, is applied to monochromatic, polychromatic and diffusive waves. Monochromatic waves may form spatially localized waves in free space, whereas polychromatic waves form non-decaying traveling evanescent modes confined to subwavelength regions in media where the frequency depends on the wave vector. We suggest an analogy between the phase compensation method and the transformation of frequencies between inertial, relativistic coordinate systems.
\end{abstract}

\pacs{Valid PACS appear here}
\maketitle

\section{Introduction}
Physicists and engineers often need to find methods for confining a field as much as possible with a minimum of knowledge of the system at hand. In the case of monochromatic waves, it has been found that phase and amplitude wave engineering is of crucial importance for subwavelength focusing of waves\cite{Goodman,Pendry,Merlin}. In the case of pulsed waves, the situation is more complex, mainly due to the fact that the superposition of many spectral components in most cases tend to broaden the focal spot. Theoretically, one may hope to be able to focus strongly confined pulses as in Ref. \cite{Sherman}, but the aperture field and medium required remains elusive. However, there are a few promising pathways, e.g., using rainbow-colored apertures which combine to a single white-colored focus\cite{Zhu} or diffractive optical elements\cite{Yero}. Well developed theories and numerical algorithms are available for studying monochromatic and polychromatic waves in focal regions\cite{Sherman,Helseth1,Veetil,Bruegge}, but they do not provide direct insight into the reverse engineering problem and may not provide a fast solution or be optimal with the minimal amount of information given. Of particular interest here is the emerging field of evanescent wave focusing, where up to now mainly monochromatic waves have been studied\cite{Merlin,Helseth,Tsukerman,Intaraprasonk,Li}. However, also for waves containing several frequency components it is necessary to have straightforward methods which allow backtracking of the signal in a manner such that the initial field resulting in a strong focus is found. In the present work we detail a method for finding the intensity distribution near the focal region using phase and amplitude compensation. In order to achieve this goal, we pinpoint a certain distribution at focus and then try to calculate the required aperture field. Moreover, we show an analogy between our compensation method and the transformation of frequencies between inertial, relativistic coordinate systems.

\section{Monochromatic waves}
The method of phase and amplitude compensation has been described in a number of studies (see e.g. Ref.\cite{Goodman,Pendry,Stamnes} and references therein), but was more recently re-introduced to study focusing of evanescent waves\cite{Merlin}. The procedure is most easily demonstrated for scalar monochromatic waves satisfying the Helmholtz equation
\begin{equation}
\left( \nabla ^{2} + k^{2} \right) E (x,z)=0 \,\,\, ,
\label{AA}
\end{equation}
where $k=\omega /v$ is the wavenumber, $\omega$ is the angular frequency and $v$ is the velocity of light in the medium under consideration.
Using the angular spectrum representation, the field can be written as
\begin{equation}
E (x,z) =\frac{1}{2\pi} \int_{-\infty}^{\infty}
E _{k} (k_{x})e^{ik_x x +ik_z z} dk_{x}\,\,\, ,
\label{BB}
\end{equation}
where the angular spectrum is given by
\begin{equation}
E_{k} (k_{x}) =\int_{-\infty}^{\infty} E (x',0)
e^{-ik_{x} x'} dx' \,\,\, ,
\label{CC}
\end{equation}
where
\begin{displaymath}
k_{z} = \left\{ \begin{array}{ll}
\sqrt{k^{2} -k_{x}^{2}} & \textrm{if $k^{2} \geq k_{x}^{2} $}\\
i\sqrt{k_{x}^{2} - k^{2}} & \textrm{if $k^{2} < k_{x}^{2} $}\\
\end{array} \right.
\label{DD}
\end{displaymath}
Here $k^{2} > k_{x}^{2}$ represent the homogenous plane
waves, whereas $k^{2} < k_{x}^{2}$ correspond to inhomogeneous or
evanescent plane waves that propagate in the $x$ direction and decay in the z - direction.

The trick is now to express the angular spectrum as
\begin{equation}
E_{k} (k_{x}) = f_{k}(k_x)e^{-ik_{z} z_0} \,\,\, ,
\label{perfect}
\end{equation}
where we must again require $E(k_x)$ to be bandlimited (i.e. nonzero only in a range $k_a \leq k_x \leq k_b$) and converges sufficiently quickly such that the solution for the field does not possess any unphysical divergency problem. The conditions for this was discussed in Ref. \cite{Helseth}, and will not be repeated here. The field at the focal point ($z=z_0$) is now given by
\begin{equation}
E(x,z_0) =\frac{1}{2\pi} \int_{-\infty}^{\infty}
f _{k} (k_{x})e^{ik_x x} dk_{x}\,\,\, ,
\label{perfectfocus}
\end{equation}
i.e. a Fourier transform of the aperture function $f(k_x)$ resulting in a finite resolution at the focal point. As an example, assume now that $f_{k}(k_x)=E_0$ for $|k_x| \leq k_a$ ($k_a \ll k$) and zero elsewhere. Then we have
\begin{equation}
E (x,z_0) =\frac{E_0}{2\pi} \int_{-k_0}^{k_0}
e^{ikx} dk =E_0\frac{sin(k_0 x)}{\pi x}\,\,\, .
\label{H}
\end{equation}
In order to find the field $E(x,0)$ we must evaluate
\begin{equation}
E(x,0) =\frac{1}{2\pi} \int_{-k_0}^{k_0}
f_{k}(k_x) e^{ik_x x -ik_z z_0} dk_{x}\,\,\, .
\label{perfectfocus2}
\end{equation}
In Ref.\cite{Helseth} it was found that in the case of paraxial waves ($k_x \ll k$) the field $E(x,0)$ is approximately a quadratic phase function, which has been described in detail in standard textbooks on optics\cite{Goodman}. Thus, in the paraxial approximation the phase compensation method directly leads us to the well-known phase profile for a lens giving the most strongly focused profile\cite{Goodman}. Recently, the phase compensation method has been used to design evanescent waves in focal regions\cite{Merlin,Helseth}. The simplest wave profile exhibiting a focal region is found by letting $f_k(k_x)=E_0$ if $k_a \leq k_x \leq k_b$ and zero elsewhere. Here $k_a \gg k$ and $E_0$ is a constant. The field can then be expressed as
\begin{equation}
E(x,z) \approx \frac{E_0}{2\pi}\left[ \frac{e^{ik_b x- k_b (z-z_0)} -e^{ik_a x- k_a (z-z_0)}}{ix -\left( z-z_0 \right)} \right] \,\,\, .
\label{perf}
\end{equation}
The intensity at the focus is therefore just $\propto sinc^{2}\left( \frac{k_b-k_a}{2}x \right) $ whereas the aperture field oscillates rapidly within an envelope such that the intensity is $\propto 1/(x^{2}+z_0^{2} )$. Notice that the aperture field is much stronger than the field in the focal region, i.e.  $E(x,0)/E(x,z_0) \sim e^{k_bz_0}$. However, we also see that the intensity distribution is more confined at the focal region than anywhere else. In Fig. \ref{f1}(a) the intensity of a monochromatic, evanescent wave is displayed with $k_a=0.1$ (arbitrary units), $k_b=1$ (arbitrary units) and $k_bz_0 =5$. The solid line shows the field in focus, whereas the dashed line shows the fields at $z=0$. This illustrates our point, namely that we may obtain a more spatially confined intensity distribution (although the magnitude is considerably smaller than that at the aperture). If we assume that $k_b \gg k_a$, a possible criterium for focusing could be obtained by requiring that the aperture intensity (at $z=z_0$) envelope at $x=z_0$, corresponding to the position with half the maximum intensity, should be wider than the focused intensity distribution of half-width $\sim \pi /k_b$. Thus, by requiring $k_b z_0 \geq \pi$, we ensure that the field is focused. At the same time, the field should not diverge at the aperture, and it is clear that $k_b z_0$ cannot be too large.

\section{Diffusive waves}
\begin{figure}
\includegraphics[width=12cm]{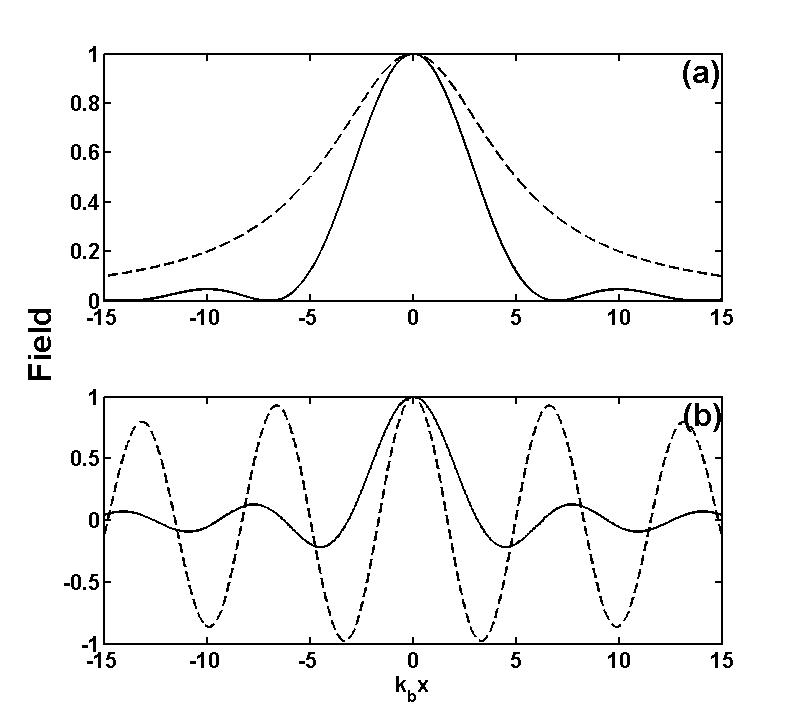}
\caption{\label{f1} In (a) the intensity of a monochromatic, evanescent waves is displayed with $k_a=0.1$ (arbitrary units), $k_b=1$ (arbitrary units) and $k_bz_0 =5$. In (b) the diffusive waves (a) with $k_0=1$ (arbitrary units), $D=1$ (arbitrary units) and $t_0=10$ are displayed.
 The solid lines show the field in focus, whereas the
dashed lines show the fields at $t=0$ for diffusive waves and $z=0$ for monochromatic evanescent waves.}
\vspace{2cm}
\end{figure}

Interestingly, the compensation method described above is not limited to designing waves in focal regions described by the Helmholtz equation. It can also be used to construct solutions to other differential equations, such as the diffusion equation, and we will here look at one example. Imagine that we generate a spatial magnetic field distribution by positioning a system of co-aligned current-carrying wires in a homogeneous medium of frequency-independent conductivity $\sigma$ and permittivity $\epsilon$. Each of the wires carry a low frequency current where the frequencies of any temporal waveform fulfills $\omega \ll \sigma /\epsilon$. In absence of sources the resulting electric (or magnetic) field can be described by the diffusion equation on the form\cite{Jackson}
\begin{equation}
D\nabla ^{2} E(x,t)= \frac{\partial E (x,t)}{\partial t} \,\,\, .
\label{Adiff}
\end{equation}
where $D=1/(\sigma \epsilon )$ is the diffusion coefficient. We now want to describe the field in the vicinity of the wires, but not including any sources such that Eq. \ref{Adiff} is valid. The field can then be written on the following general form
\begin{equation}
E (x,t) = \frac{1}{2\pi} \int_{-\infty}^{\infty}
E_{k}(k) e^{ikx -Dk^{2}t } dk  \,\,\, ,
\label{HHH}
\end{equation}
where $E_{k} (k)$ is the fourier amplitude. Usually, a solution of eq. \ref{HHH} will represent a diffusion process where the field delocalizes with increasing time\cite{Jackson}. However, we are here interested in finding a field $E(x,0)$ with an optimally localized field $E(x,t_0)$ ($t_0>0$). To see how one may design such an initial field, let us assume that $E_k(k)=E_0\exp(Dk^{2}t_0)$ is a constant for $|k| \leq k_0$ and zero for $|k| \geq k_0$. Thus, the field is given by
\begin{equation}
E (x,t) = \frac{E_0}{2\pi} \int_{-k_0}^{k_0}
e^{ikx -Dk^{2}(t-t_0)} dk \,\,\, ,
\label{HH1}
\end{equation}
The real value of eq. \ref{HH1} at $t=0$ is given by
\begin{equation}
E (x,0) = \frac{E_0}{2\pi} \int_{-k_0}^{k_0}
cos(kx) e^{Dk^{2}t_0} dk\,\,\, ,
\label{HH2}
\end{equation}
whereas at $t=t_0$ it is
\begin{equation}
E (x,t_0) =\frac{E_0}{2\pi} \int_{-k_0}^{k_0}
e^{ikx} dk =E_0\frac{sin(k_0 x)}{\pi x}\,\,\, ,
\label{H}
\end{equation}
By selecting $Dt_0$ such that $Dk_0^{2}t_0 \gg 1$, we see that the leading contribution to $E(x,0) \sim cos(k_0x) e^{Dk_0^{2}t_0}$, which is an oscillatory function with period $\Delta x= 2\pi/k_0$. Thus, the field at $t=0$ is strongly delocalized, but with sharp local fluctuations, as seen in Fig. \ref{f1} b). On the other hand, $E(x,t_0)$ is strongly localized with a central lobe $2\pi/k_0$, but is a factor $e^{Dk_0^{2}t_0}$ smaller than the field at $t=0$. A sharp localization is therefore obtained at the expense of an exponential reduction in field amplitude. This is rather similar to the behavior we observe for monochromatic evanescent waves, although the details are different.
In addition to describing electromagnetic fields, the diffusion equation may also describe concentration or temperature (but now with a different diffusion coefficient). In those cases it should be mentioned that we do not consider absolute temperature or concentrations (as required by our boundary conditions), but rather modulations about an equilibrium. Thus, the negative values seen in Fig. \ref{f1} b) do not mean absolute negative temperature and concentration, they are just signatures of the variations about the mean temperature or concentration seen when designing an initial profile required to bring the diffusive waves to a focus.

\section{Polychromatic waves}
An interesting question is whether the method above can be applied directly to polychromatic waves (e.g. pulsed waves) satisfying the wave equation
\begin{equation}
\nabla ^{2} E (x,z,t)= \frac{1}{c^{2}} \frac{\partial ^{2} E(x,z,t)}{\partial t^{2}}
\label{timeA}
\end{equation}
In the angular spectrum representation, the field can be expressed as
\begin{equation}
E(x,y,z,t) = \int_{0}^{\infty}
\int_{-\infty}^{\infty} E_{k} (k_{x}, \omega)
e^{ik_x x +ik_z z -i\omega (k_x) t} dk_{x}d\omega  \,\,\, ,
\label{B}
\end{equation}
where $k=\omega(k_x)/c$ and
\begin{displaymath}
k_{z} = \left\{ \begin{array}{ll}
\sqrt{k^{2} -k_{x}^{2}} & \textrm{if $k^{2} \geq k_{x}^{2} $}\\
i\sqrt{k_{x}^{2} - k^{2}} & \textrm{if $k^{2} < k_{x}^{2} $}\\
\end{array} \right.
\label{C}
\end{displaymath}

In order to bring the polychromatic waves to focus we require cross-compensation of the phase, which amounts to setting $k_z z_1 -\omega (k_x)t_1 =0$, i.e. the spatial part of the phase is compensated by the temporal part at $(x,z_1,t_1)$. Such a requirement can only be fulfilled if $\omega$ depends on the spatial frequencies (i.e. direction of each plane wave), where the angular frequency is given by $\omega (k_x) =i\gamma (v_1) |k_x| v_1$. Here $\gamma (v_1) =1/\sqrt{1-(v_1/c)^{2}}$ and $v_1=z_1/t_1$.
The fact that the frequency of the spatial wave vector depends on the direction was utilized in Ref. \cite{Zhu} to combine a rainbow spectrum to a single, focused spot. Conceptually the idea presented here is somewhat similar, although we use an entirely different approach to achieve the goal. Note that we must distinguish between the two cases $v_1 \geq c$ and $v_1 \leq c$.

The case $v_1 \geq c$ results in superluminal, localized waveforms which have been discussed extensively in the literature\cite{Besieris,Ciattoni}. Here $\omega (k_x)$ is real and given by $\omega (k_x)=\alpha k_x$ ($\alpha = v_1/\sqrt{(v_1 /c)^{2}-1}$). The scalar field is then given by
\begin{equation}
E(x,z,t) = \int_{0}^{\infty}
\int_{-\infty}^{\infty} E_{k} (k_{x},\omega)
e^{ik_x x + i|k_x| \alpha \left( t-\frac{z}{v_1} \right)} dk_{x}d\omega\,\,\, .
\label{E}
\end{equation}
A similar representation was found in ref. \cite{Ciattoni} describing one-dimensional nondiffracting pulses, but here we have arrived at Eq. \ref{E} from an entirely different perspective based on requirement of specific field profile resulting from phase compensation.

Of greater interest in the current study is the case $v_1 \leq c$, since that represents an evanescent wave solution similar to that seen in the previous section. Now $\alpha$ is purely imaginary, such that
\begin{equation}
E(x,z,t) = \int_{0}^{\infty}
\int_{-\infty}^{\infty} E_{k} (k_{x},\omega)
e^{ik_x x  -|k_x| \gamma \left( t-\frac{z}{v_1} \right)} dk_{x}d\omega \,\,\, .
\label{F1}
\end{equation}
As an example, we may approximate the angular spectrum as $E(k_x,\omega)=E_0$ for $k_a \leq k_x \leq k_b\, , \, \omega _1 \leq \omega \leq \omega _2$ and zero elsewhere. Here we also assume that
$k_a \gg k$, such that only evanescent waves are excited with $k_z\approx i|k_x|$ and $\omega (k_x) \approx i|k_x| v_1$. This evanescent-wave approximation therefore requires $v_1 \ll c$, and
the field will be
\begin{equation}
E(x,z,t) \propto  \left[ \frac{e^{ik_b x- k_b (z-v_1 t)} -e^{ik_a x- k_a (z-v_1 t)}}{ix -\left( z-v_1 t \right)} \right] \,\,\, .
\label{evantime}
\end{equation}
Note that eq. \ref{evantime} is identical to eq.\ref{perf} if we set $z_0=v_1 t$, and is therefore just a translation of the evanescent wave along the optical axis with the intensity distribution similar to that of fig. \ref{f1}. Since we require $k_b \gg k$,
which is fulfilled if we, as an example, set $k_b=10k=20\pi /\lambda$, the half-width of the intensity distribution at the traveling focal line
is narrower than $\sim \pi/k_b=\lambda /20$. The intensity at $z=0$ grows as $I(x,0,t>0)\propto \exp(k_b v_1 t)/(x^{2} + (v_1t)^{2})$. However, when $t \gg 1/k_b v \sim t_1$ our approximate theory for evanescent waves does not longer hold in the vicinity of $z=0$ (this follows from the considerations above; see also Refs. \cite{Stamnes,Helseth} for a discussion about diverging solutions). The fact that
the evanescent wave does not change at the focal line $z_1=v_1 t_1$ is surprising, given the condition $E(x,0,0)=E(x,z_1,t_1)$ above. However, we also note that $E(x,0,t_1) \gg E(x,z_1,t_1)$ such that the increasing energy of the field at the aperture is used to keep the field at the focal line $z_1=v_1 t_1$ unchanged in magnitude as it propagates outwards. 
The technical implementation of focusing of polychromatic waves is in effect similar to that of monochromatic waves, but with two new important features: a) The focusing must take place in a spatially dispersive media, and b) The field at the aperture must grow with time (within the approximation given above).

An interesting analogy occurs if one compares the problem of designing an aperture field that generate optimal focus with that of transformation of frequencies between inertial, relativistic coordinate systems\cite{Einstein}. To see this, consider a plane wave of the form $\exp(k_z z_1 -\omega (k_x)t_1 + \phi (k_x))$. Any wavepacket is a weighted sum over such plane waves in spatial and frequency coordinates. In order to bring the wave packet to focus we require that every spectral component is at focus a certain distance $z_1$ from the aperture at a given time $t_1$. This can be done by setting $k_z z_1 -\omega (k_x)t_1 + \phi (k_x)=0$, such that the phase is exactly compensated at $(z_1,t_1)$. It should be noted, as seen above, that such a requirement may give rise to strong localization at other positions as well. In any case, such a cross-compensation requirement can only be fulfilled if $\omega$ depends on the spatial frequencies (i.e. direction of each plane wave), where the frequency is given by
\begin{equation}
\omega (k_x) = -\frac{\phi (k_x)/t_1}{\left( \frac{v_1}{c}\right) ^{2} -1} \pm \frac{\phi (k_x)/t_1}{\left( \frac{v_1}{c}\right) ^{2} -1} \sqrt{1+\left[ \left( \frac{v_1}{c}\right) ^{2} -1 \right]
\left[ \left( \frac{k_x z_1}{ \phi (k_x)} \right) ^{2} -1 \right]} \,\,\, .
\label{timeC}
\end{equation}
Now consider the problem of transforming plane waves between inertial, relativistic coordinate systems. That is, consider an observer moving at a speed $v$ relative to a fixed frame. According to Einstein's special theory of relativity, the moving observer will detect a frequency $\omega '$ given by $\omega ' =\gamma (v) (\omega -k_z v)$, where $\gamma (v) =1/\sqrt{1-(v/c)^{2}}$\cite{Einstein}. From the above it may be inferred that this expression corresponds to $k_z z_1 -\omega (k_x)t_1 +i\phi (k_x) =0$ if we make the associations $\phi(k_x)/t_1 \rightarrow \omega'/\gamma (v_1)$ and $v_1 \rightarrow v$, i.e. the phase factor $\phi (k_x)$ must be associated with the frequency measured in the moving frame. In the special case considered in this study $\phi (k_x)=0$, which corresponds to $\omega ' =0$. For evanescent waves we then immediately find the spatial dispersion relationship given above. For propagating waves we may set $k_z=(\omega/c)cos\theta$, where $\theta$ is the real angle at which a specific plane wave makes with the direction of motion. Then we must have $cos\theta =c/v$ and therefore $v\geq c$, in agreement with the observations made above.

\section{Conclusion}
In conclusion, we have suggested a compensation method for designing aperture fields giving rise to strongly confined waves. The idea is to first try to compensate the phase or amplitude such that only transverse spatial frequencies are left in the angular spectrum representation at a given region in space, thus allowing a strongly focused wave to form here. Next we calculate the aperture field required to obtain such a phase or amplitude compensation. The method has been employed to study diffusive, monochromatic and polychromatic waves, and shown to give new insight into the problems at hand. The method here can probably also be applied to other wave systems where phase or amplitude compensation is beneficial.

\newpage

\newpage

\end{document}